# Diode-laser-pumped continuous-wave mid-infrared optical parametric oscillator

Ville Ulvila,[1,2] Markku Vainio[1,3,*]

[1]*Molecular Science, Department of Chemistry, University of Helsinki, Finland*

[2] *VTT Technical Research Centre of Finland Ltd., Finland*

[3]*Laboratory of Photonics, Tampere University of Technology, Tampere, Finland*
**Corresponding author: markku.vainio@helsinki.fi*

Abstract: We report a singly resonant continuous-wave optical parametric oscillator (SRO), which is pumped by a semiconductor laser system and operates in the mid-infrared region (2.9 to 3.6 µm). Fast tuning of the mid-infrared output wavelength by more than 200 nm is demonstrated by scanning the pump laser wavelength through only 1.25 nm, without any need to adjust the SRO settings. The exceptionally large tuning range is obtained by choosing a pump wavelength (780 nm) that corresponds to a turning point of the group velocity mismatch between the phase-matched signal and idler waves of the SRO.

Continuous-wave singly resonant optical parametric oscillators (SROs) are coherent light sources that provide a rare combination of large wavelength tuning range, narrow linewidth, and high output power [1-5]. These sources are particularly useful in the 3 to 5 µm wavelength range, which is one of the important fingerprint regions used in molecular spectroscopy [6]. While the CW SROs based on quasi-phase-matched (QPM) nonlinear crystals are established tools in laboratory experiments, these devices are generally too bulky and expensive for field applications, such as for trace gas monitoring. The remaining limitations are mainly due to the need for an expensive high-power pump laser, but also due to the rather slow SRO wavelength tuning mechanisms, which typically require adjustment of the nonlinear crystal temperature and/or QPM period [1, 3, 6]. Here, we demonstrate an approach to overcome these limitations, in order to realize a field-compatible mid-infrared CW SRO with fast and simple wavelength tuning.

Our solution is based on semiconductor laser pumping, which has previously been demonstrated with near-infrared SROs [7]. The development of high quality mirror coatings and QPM crystals now makes it possible to achieve diode-laser pumped CW SRO operation also in the important 3 to 5 µm region using commercially available components. More importantly, diode-laser pumping makes it possible to extend the SRO tuning range achievable by a simple field-compatible pump tuning method. This is illustrated in Fig. 1, which shows the phase-matched SRO output wavelengths as a function of pump wavelength for various QPM settings. The tuning curves of Fig. 1 are calculated for a typical MgO-doped periodically poled $LiNbO_3$ (MgO:PPLN) SRO for a fixed crystal temperature. For the most frequently used pump wavelength, 1064 nm, the idler wavelength depends on the pump wavelength only weakly, except for the region close to signal-idler degeneracy, where useful CW SRO operation is impeded due to spectral instabilities. Although pure pump-tuning of SRO by >150 nm has been reported by utilizing a broadly tunable ~1064 nm laser [8] and a large pump acceptance bandwidth [9], wide SRO tuning typically requires adjustment of the crystal temperature and/or QPM period, both of which are slow. The approximately 760 to 860 nm region, on the other hand, allows tuning of the mid-infrared idler wavelength over hundreds of nanometers by moderate pump wavelength tuning. This region, where the group velocity mismatch changes sign, is also useful for broadband parametric generation and amplification [5, 10].

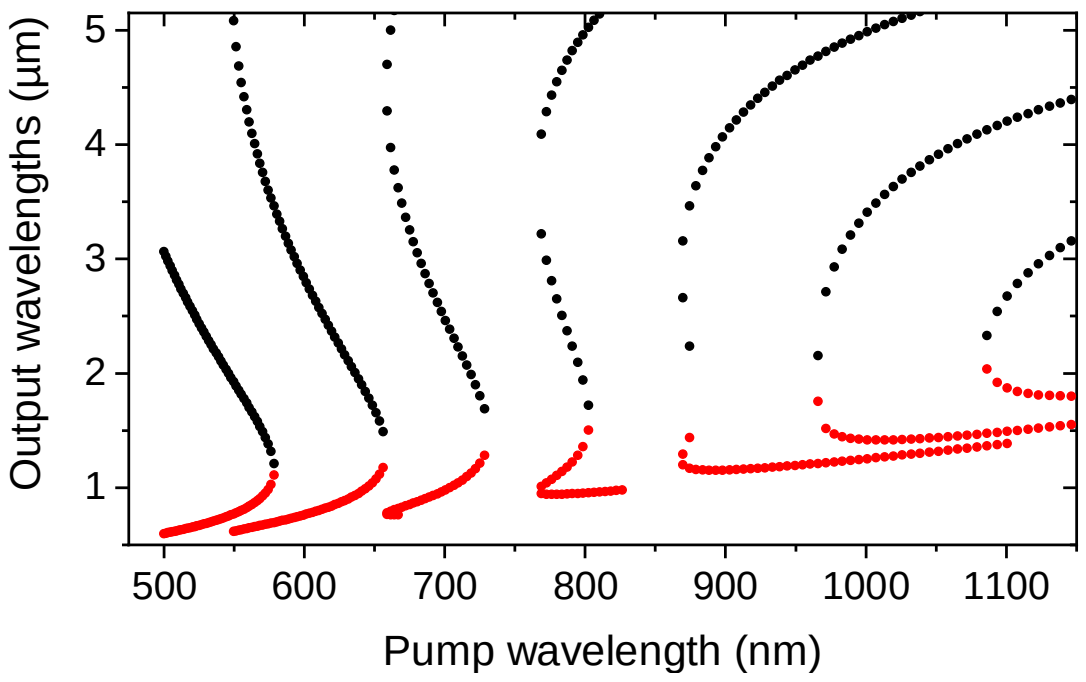

Fig. 1. Calculated phase-matched signal and idler wavelengths of a MgO:PPLN-based SRO for seven different QPM periods, ranging from 9 to 33 µm (left to right) in increments of 4 µm [11]. The crystal temperature is 320 K and all interacting waves are *e*-polarized.

The 780 nm region is easily accessible with narrow-linewidth distributed-feedback (DFB) diode lasers and tapered semiconductor amplifiers suitable for cost-efficient CW SRO pumping, as we demonstrate in this Letter. It is also worth pointing out that in comparison with 1064 nm, the shorter 780 nm pump wavelength reduces the SRO oscillation threshold [5] without significantly increasing the photorefractive damage, which requires special attention when pumping at 532 nm [12].

A prominent feature of the tuning curves of Fig. 1 is that the sign of change in the phase-matched idler wavelength vs. pump wavelength tuning changes when the pump wavelength is varied from visible to near infrared. To understand this behavior, let us consider phase-matched SRO operation: $\Delta k = k_p - k_s - k_i - 2\pi/\Lambda = 0$, where $k_j = n_j\omega_j/c$, with sub-index j referring to either pump (p), signal (s), or idler (i). Here, $c$ is the speed of light, $n_j$ is the refractive index, $\omega_j$ is the angular frequency, and $\Lambda$ is the QPM period of the crystal. If the pump frequency is changed, for a small frequency detuning $\delta\omega_p$ the resulting wave-vector mismatch can be approximated to the first order as [9, 13]

$$\Delta k = \delta\Delta k \approx \frac{\partial k_p}{\partial\omega_p}\delta\omega_p - \frac{\partial k_s}{\partial\omega_s}\delta\omega_s - \frac{\partial k_i}{\partial\omega_i}\delta\omega_i = \frac{\partial k_p}{\partial\omega_p}\delta\omega_p - \frac{\partial k_s}{\partial\omega_s}(\delta\omega_p - \delta\omega_i) - \frac{\partial k_i}{\partial\omega_i}\delta\omega_i = \delta\omega_p\left(\frac{\partial k_p}{\partial\omega_p} - \frac{\partial k_s}{\partial\omega_s}\right) - \delta\omega_i\left(\frac{\partial k_i}{\partial\omega_i} - \frac{\partial k_s}{\partial\omega_s}\right) = \delta v_{ps}\delta\omega_p - \delta v_{is}\delta\omega_i, \quad (1)$$

where $\delta\omega_s$ and $\delta\omega_i$ are the resulting small changes in the signal and idler frequencies, respectively. Photon energy is conserved in the SRO process, thus $\delta\omega_p = \delta\omega_s + \delta\omega_i$. The group-velocity mismatch (GVM) terms are defined as $\delta v_{ps} = v_{g,p}^{-1} - v_{g,s}^{-1}$ and $\delta v_{is} = v_{g,i}^{-1} - v_{g,s}^{-1}$, where $v_{g,j} = \frac{\partial\omega_j}{\partial k_j}$ is the group velocity. Recall that the group index is defined as $n_{g,j} = \frac{c}{v_{g,j}}$, so we can also write $\delta v_{ps} = (n_{g,p} - n_{g,s})/c$ and $\delta v_{is} = (n_{g,i} - n_{g,s})/c$.

In order to maintain parametric oscillation while tuning the pump frequency, one should retain $\Delta k \approx 0$. Using Eq. (1), this leads to

$$\frac{\delta\omega_i}{\delta\omega_p} \approx \frac{\delta v_{ps}}{\delta v_{is}}, \quad (2)$$

which implies that the sign and slope of idler frequency change vs. pump detuning depend on the ratio of the two GVM terms. In practice, the sign of $\delta\omega_i/\delta\omega_p$ is determined by the sign of $\delta v_{is}$, because $\delta v_{ps}$ is always positive ($n_{g,p} > n_{g,s}$) for all cases presented in Fig. 1. This can be seen from Fig. 2, where we have plotted the group index of MgO:PPLN as a function of wavelength [14]. The figure shows examples of typical pump, signal, and idler group index values of the SRO of this work. As another example, the most common SRO configuration, which is pumped at 1064 nm, produces idler and signal wavelengths of approximately 3 and 1.65 µm, respectively. From Fig. 2 we get $n_{g,p} > n_{g,i} > n_{g,s}$ for this particular configuration.

For short pump wavelengths, such as 532 nm, the GVM term $\delta v_{is}$ is always negative. Thus, the phase-matched tuning curve (Fig. 1) is monotonous with a negative slope, $\delta\omega_i/\delta\omega_p < 0$. For long pump wavelengths, such as 1064 nm, the slope is of opposite sign, $\delta\omega_i/\delta\omega_p > 0$. The GVM term $\delta v_{is}$ changes sign for a phase-matched signal-idler pair that corresponds to a pump wavelength of about 780 nm, and phase matching can be obtained for both $\delta v_{is} < 0$ and $\delta v_{is} > 0$. This leads to a tuning curve that has two branches with opposite tuning directions: the "inner" branch ($\delta v_{is} < 0$) and the "outer" branch ($\delta v_{is} > 0$), as illustrated in Fig. 1 and in more detail in Fig. 3a. Either of the two tuning branches can be accessed, as we have previously demonstrated using a Ti:sapphire-laser pumped CW SRO [2]. Here, we have designed our SRO to work on the inner branch, which gives a broad wavelength tuning range in the 3 µm region that is particularly interesting for molecular spectroscopy applications [6].

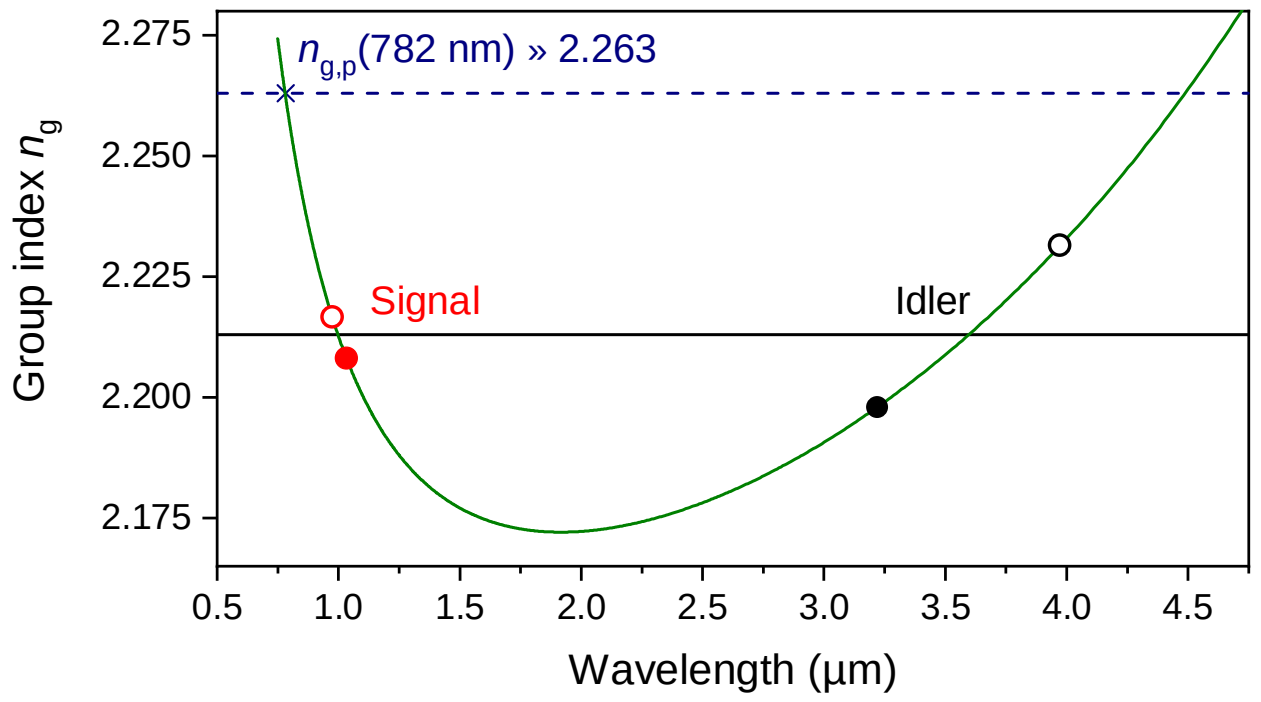

Fig. 2. Group index of MgO:PPLN at 303 K temperature. The open circles show an example of a phase-matched signal-idler pair of a 782-nm pumped SRO of the "outer" tuning branch of Fig. 3a ($n_{g,i} > n_{g,s}$). The solid dots exemplify a signal-idler pair of the "inner" branch, where $n_{g,i} < n_{g,s}$. The solid line indicates the GVM turning point $n_{g,i} = n_{g,s}$.

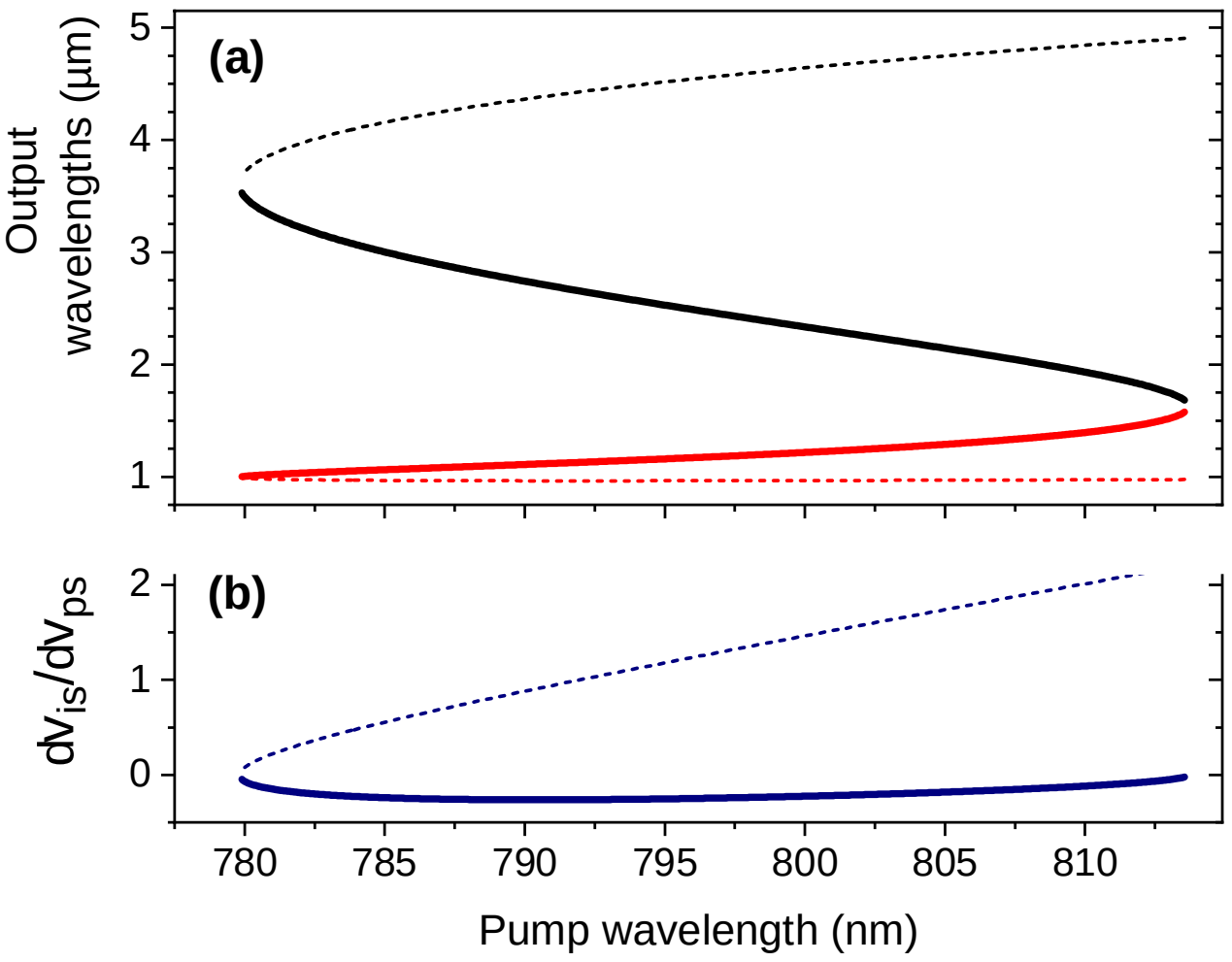

Fig. 3. (a) Phase-matched signal (red) and idler (black) wavelengths of a MgO:PPLN SRO computed using the Sellmeir data of [14]. MgO:PPLN temperature and QPM period are 303 K and 21.5 µm, respectively. (b) The GVM ratios of the phase-matched wavelengths. In both (a) and (b) the solid (dashed) lines correspond to $\delta v_{is} < 0$ ($\delta v_{is} > 0$).

Figure 3b shows the GVM ratios $\delta v_{is}/\delta v_{ps}$ that correspond to the phase-matched signal-idler pairs of Fig. 3a. In addition to being useful in explaining the direction and slope of the idler tuning vs. pump tuning, this ratio predicts the susceptibility of the SRO to multimode operation [13]. Single-mode operation, which is crucial for molecular spectroscopy applications, can only be achieved if the SRO is modulationally stable, *i.e.* stable against small perturbations. The instability threshold $N_{th}$ (defined as the ratio of the pump power at instability threshold to the pump power at SRO oscillation threshold) depends on $\delta v_{is}/\delta v_{ps}$ and on the group velocity dispersion. For the inner tuning branch of Fig. 3, $\delta v_{is}/\delta v_{ps}$ is between 0 and –0.25, which leads to a $N_{th} < 2$, see Fig. 4 of [13]. In practice, this means that an intracavity etalon or another frequency selective cavity component is needed to suppress the instability. For the outer tuning branch $\delta\upsilon_{is}/\delta\upsilon_{ps} > 0$, which provides $N_{th} > 3$. In this case, stable single-mode operation can often be obtained even without an intracavity etalon.

The experimental setup used in this work is outlined in Fig. 4. Based on the above considerations, we have chosen 780 nm wavelength and a master-oscillator-power-amplifier configuration for the SRO pump laser. The master oscillator is a DFB diode laser (Eagleyard EYP-DFB-0780-00040-1500-BFX02-0002), and the amplifier is a 2 W tapered semiconductor amplifier (Thorlabs TPA780P20). The SRO cavity is a standard travelling-wave bow-tie cavity, designed to resonate the signal wavelength. A 5-cm long MgO:PPLN crystal with AR-coatings (770-860 (R<2%) 1000-1180 (R<0.25%) 2500-4500 (R<10%)) is placed between the two concave mirrors of the cavity, such that both the pump and signal beam waist sizes in the crystal correspond to a focusing parameter of approximately 2. In order to keep the setup simple, astigmatism of the pump beam was not compensated. Both concave mirrors of the SRO have a radius of curvature of 75 mm, which leads to a small cavity footprint (*c.a.* 80 mm x 150 mm). All cavity mirrors have a high reflectivity of >99.8% in the signal wavelength range (1000 – 1070 nm), and high transmission for the pump (780 nm) and idler (2.9 – 3.6 µm) wavelengths. A 400-µm thick uncoated YAG etalon was placed inside the SRO cavity to suppress multimode operation and mode hops, as discussed above.

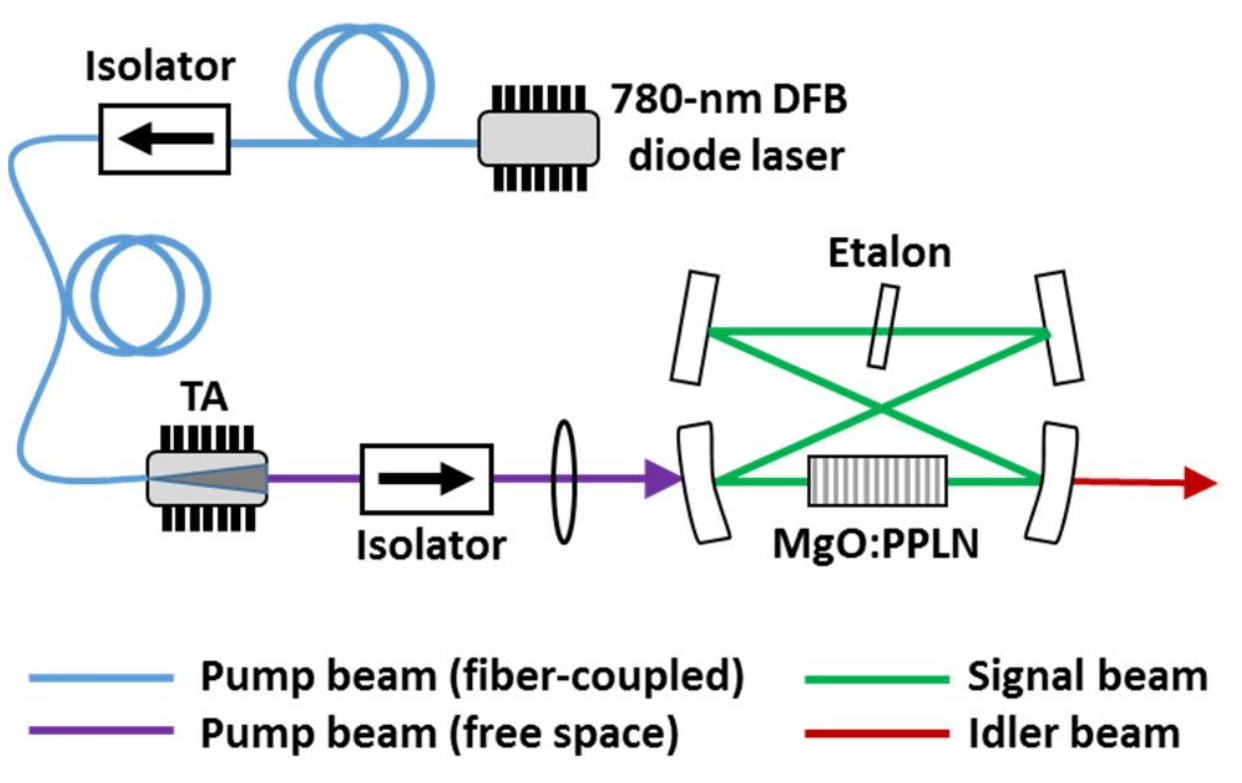

Fig. 4. Schematic of the CW SRO layout. TA = tapered amplifier.

The total tuning range of the SRO idler wavelength is from 2.9 to 3.6 µm, which can be covered using two different QPM periods (21.25 and 21.50 µm) and by varying the crystal temperature between 293 and 353 K. The threshold pump power, as measured between the input coupler and the crystal, is between 500 and 700 mW depending on the wavelength. The maximum CW mid-infrared power at the SRO output is c.a. 100 mW before losses caused by a dichroic mirror that was used to filter out the pump residue. This corresponds to ~75 % pump depletion and was measured with the maximum pump power of 1.2 W that was available at the crystal input. The linewidth of the mid-infrared idler beam is ~ 10 MHz, which was measured by direct beat-frequency comparison against a fully stabilized mid-infrared frequency comb [15]. Long-term stability of the idler frequency is exemplified in Fig. 5a. The measurement was in short time scales limited by the resolution of the wavelength meter, as illustrated by the inset, which shows a detail of the long recording.

In order to demonstrate a wide mid-infrared wavelength tuning by pump laser tuning only, the MgO:PPLN temperature and QPM period were kept constant (303 K and 21.5 µm, respectively) and the pump laser wavelength was scanned by changing its temperature from 289 to 312 K. This corresponds to a 1.25 nm mode-hop-free change in the DFB laser wavelength, from 780.00 to 781.25 nm. The resulting scan of the mid-infrared idler wavelength over >200 nm is illustrated in Fig. 5b and is in agreement with the theoretical tuning curve of Fig. 3a. A slow pump scan speed was used here, in order to be able to record the idler wavelength with a wavelength meter. During the scanning, SRO mode hops occur in rather predictable steps of typically *c.a.* 200 GHz or 400 GHz, both consistent with the calculated 204 GHz free spectral range of the SRO intracavity etalon. Figure 5c shows a detail of the slow scan, revealing such mode hops.

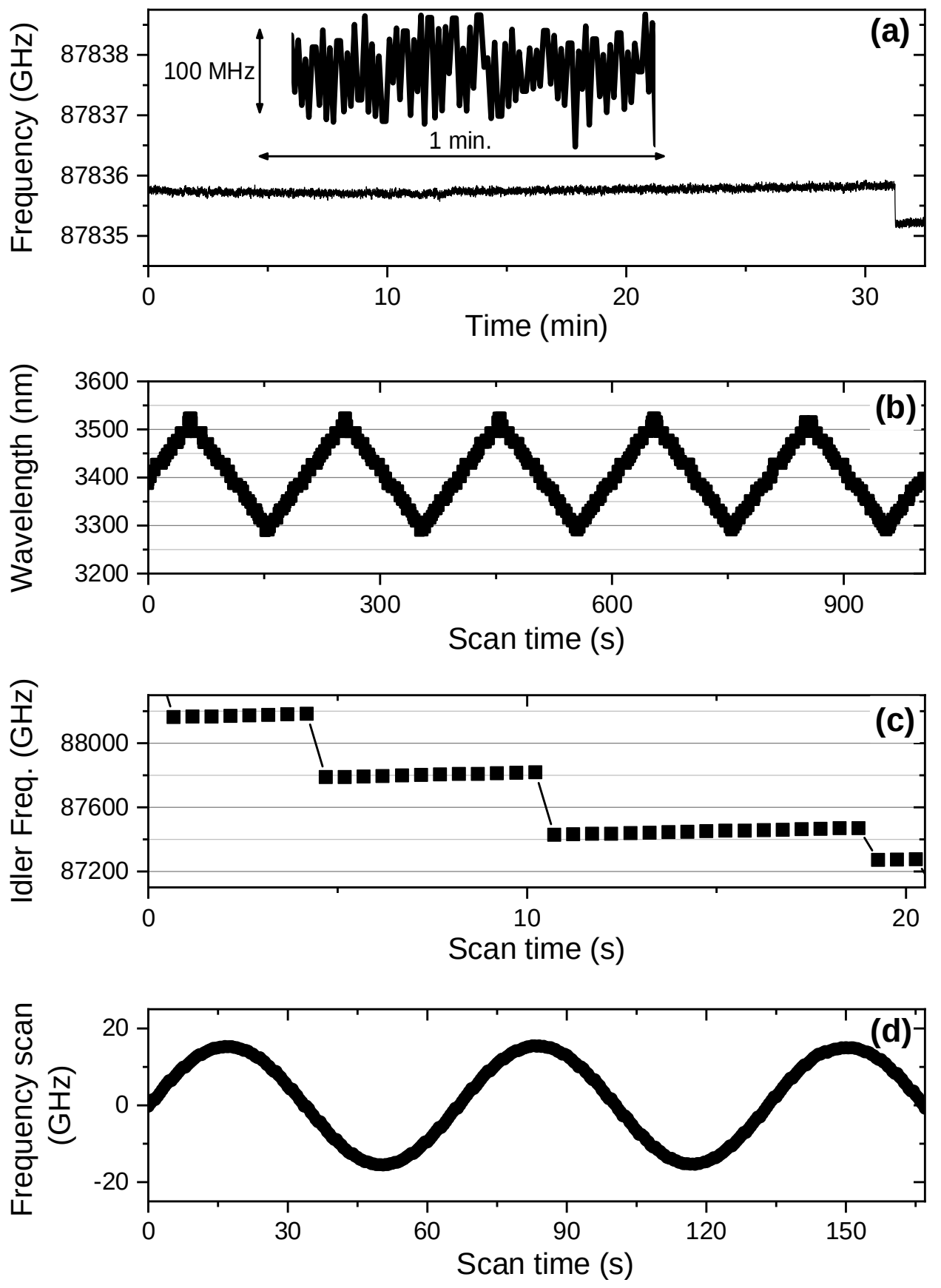

Fig. 5. (a) Idler frequency of the free-running SRO, with the inset showing a detail. (b) Tuning of the idler wavelength l i over > 200 nm performed by scanning the pump laser wavelength between 780.00 and 781.25 nm. (c) Detail of (b), showing SRO etalon mode hops. (d) Mode-hop-free idler frequency scan obtained by a 30 GHz pump laser tuning.

Mode-hop-free regions between the large etalon mode hops are typically between 10 and 40 GHz. An example of a repetitive mode-hop-free scan at one of these points is presented in panel d of Fig. 5. Note that within these mode-hop-free regions the idler frequency scan directly follows the pump frequency scan ($d\nu_i \approx d\nu_p$), since $\nu_p = \nu_s + \nu_i$ and the signal frequency is fixed. This also explains why the scan direction within the mode-hop-free regions is opposite to the global scan direction (Fig. 5c) determined by the phase-matching condition, and hence by the GVM term $\delta v_{is}$.

In order to verify the fast scanning capability of the SRO, the DFB laser frequency was tuned by changing the laser current. The change in the mid-infrared output frequency was analysed using a solid Fabry-Perot etalon made of germanium, and the power transmitted through the etalon was recorded with a fast mercury-cadmium-telluride detector. The result of such a measurement is given in Fig. 6, which shows a 26 GHz mode-hop-free scan at a scan speed of about 5.2 THz/s.

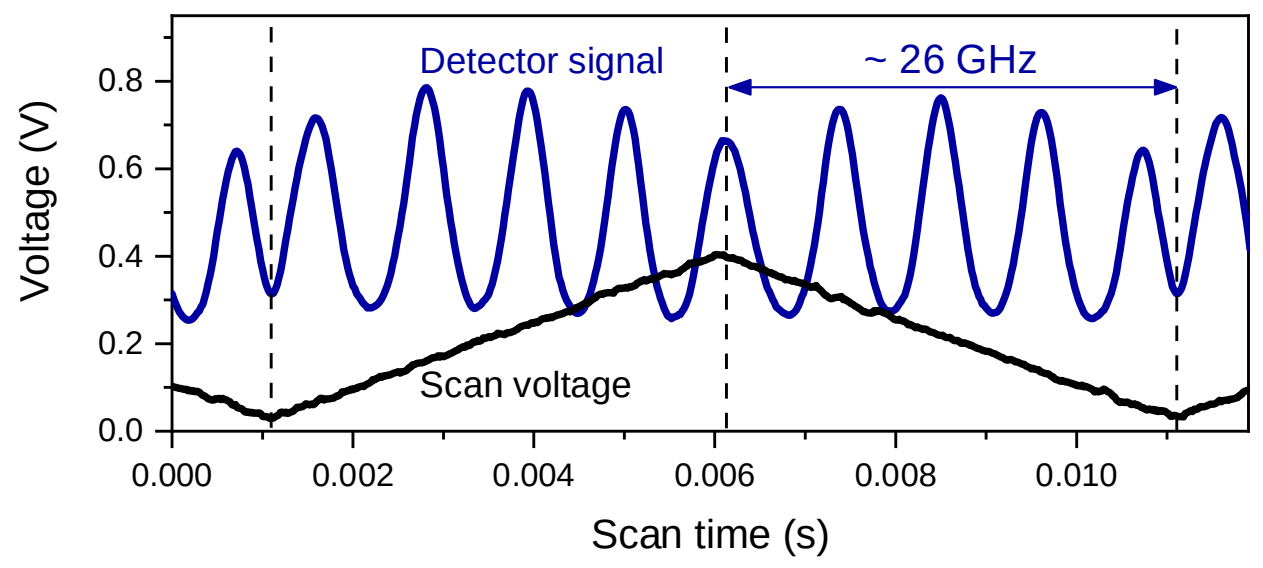

Fig. 6. Fast scan of the mid-infrared idler frequency performed by scanning the DFB pump laser current. The black triangular trace shows the scan control signal, and the blue trace is the idler power transmission through a solid Ge etalon used as an optical spectrum analyser. The free spectral range of the etalon is 6.03 GHz.

In conclusion, we have experimentally demonstrated a simple diode-laser pumped continuous-wave singly resonant optical parametric oscillator (CW SRO) that operates in the important 3-µm molecular fingerprint region. The compact and robust low-cost pump laser makes the SRO field compatible, and the output wavelength of the SRO can be rapidly tuned over 200 nm by just 1.25 nm wavelength tuning of the pump laser. Owing to GVM-optimized pump wavelength, such a large mid-infrared tuning range was obtained without adjusting any other parameters of the SRO, such as crystal temperature or QPM period. In future, even much larger mid-infrared tuning range is possible to achieve by replacing the DFB diode laser with a more widely tunable external-cavity diode laser (ECDL). As an example, 2.5 to 3.5 µm range can be covered with a single scan with the pump-tuning method using an ECDL that is tunable between 780 and 795 nm, see Fig. 3a.

The pump tuning produces large (~200 to 400 GHz) SRO mode hops attributed to the SRO intracavity etalon. Between these etalon mode hops, uninterrupted idler tuning of a few tens of GHz is obtained. The wavelengths between the etalon mode hops can be accessed by standard means, such as by varying the etalon angle – a method that is fast and can be automated [1, 16]. An interesting possibility is to synchronize the intracavity etalon scanning with pump laser scanning, such that the frequency interval of SRO mode hops is small and allows for rapid and predictable high-resolution mode-hop scanning [17].

Funding. Tekes (4218/31/2016).

Acknowledgment. We thank Mr. Juho Karhu for assistance in the SRO linewidth measurements.

## References

1. A. K. Y. Ngai, S. T. Persijn, G. von Basum, and F. J. M. Harren, "Automatically tunable continuous-wave optical parametric oscillator for high-resolution spectroscopy and sensitive trace-gas detection," Applied Physics B 85, 173-180 (2006).

2. M. Siltanen, M. Vainio, and L. Halonen, "Pump-tunable continuous-wave singly resonant optical parametric oscillator from 2.5 to 4.4 µm," Opt. Express 18, 14087-14092 (2010).

3. M. Vainio, J. Peltola, S. Persijn, F. J. M. Harren, and L. Halonen, "Singly resonant cw OPO with simple wavelength tuning," Opt. Express 16, 11141-11146 (2008).

4. A. Henderson, and R. Stafford, "Low threshold, singly-resonant CW OPO pumped by an all-fiber pump source," Opt. Express 14, 767-772 (2006).

5. L. E. Myers, R. C. Eckardt, M. M. Fejer, R. L. Byer, W. R. Bosenberg, and J. W. Pierce, "Quasi-phase-matched optical parametric oscillators in bulk periodically poled LiNbO3," J. Opt. Soc. Am. B 12, 2102-2116 (1995).

6. M. Vainio, and L. Halonen, "Mid-infrared optical parametric oscillators and frequency combs for molecular spectroscopy," Physical Chemistry Chemical Physics 18, 4266-4294 (2016).

7. M. E. Klein, C. K. Laue, D. H. Lee, K. J. Boller, and R. Wallenstein, "Diode-pumped singly resonant continuous-wave optical parametric oscillator with wide continuous tuning of the near-infrared idler wave," Opt. Lett. 25, 490-492 (2000).

8. J. Courtois, R. Bouchendira, M. Cadoret, I. Ricciardi, S. Mosca, M. De Rosa, P. De Natale, and J.-J. Zondy, "High-speed multi-THz-range mode-hop-free tunable mid-IR laser spectrometer," Opt. Lett. 38, 1972-1974 (2013).

9. R. Das, S. C. Kumar, G. K. Samanta, and M. Ebrahim-Zadeh, "Broadband, high-power, continuous-wave, mid-infrared source using extended phase-matching bandwidth in MgO:PPLN," Opt. Lett. 34, 3836-3838 (2009).

10. M. Tiihonen, V. Pasiskevicius, A. Fragemann, C. Canalias, and F. Laurell, "Ultrabroad gain in an optical parametric generator with periodically poled KTiOPO4," Applied Physics B 85, 73-77 (2006).

11. A. Smith, "SNLO software package," www.as-photonics.com (2017).

12. G. K. Samanta, G. R. Fayaz, Z. Sun, and M. Ebrahim-Zadeh, "High-power, continuous-wave, singly resonant optical parametric oscillator based on MgO:sPPLT," Opt. Lett. 32, 400-402 (2007).

13. C. R. Phillips, and M. M. Fejer, "Stability of the singly resonant optical parametric oscillator," J. Opt. Soc. Am. B 27, 2687-2699 (2010).

14.O. Gayer, Z. Sacks, E. Galun, and A. Arie, "Temperature and wavelength dependent refractive index equations for MgO-doped congruent and stoichiometric LiNbO3," Applied Physics B 91, 343-348 (2008).

15. M. Vainio, and J. Karhu, "Fully stabilized mid-infrared frequency comb for high-precision molecular spectroscopy," Opt. Express 25, 4190-4200 (2017).

16. A. M. Morrison, T. Liang, and G. E. Douberly, "Automation of an “Aculight” continuous-wave optical parametric oscillator," Review of Scientific Instruments 84, 013102 (2013).

17. S. E. Bisson, K. M. Armstrong, T. J. Kulp, and M. Hartings, "Broadly tunable, mode-hop-tuned cw optical parametric oscillator based on periodically poled lithium niobate," Appl. Opt. 40, 6049-6055 (2001).